\documentclass[pra,twocolumn,superscriptaddress,showpacs,longbibliography]{revtex4}

\usepackage{graphicx}
\usepackage{latexsym}
\usepackage{amsmath}
\usepackage{graphics}
\usepackage{amssymb}
\usepackage{layout}
\usepackage{verbatim}
\usepackage{amsfonts,epsfig,dsfont}
\usepackage{natbib}
\usepackage{hyperref}
\usepackage{color}

\newcommand{\unit}{\mathbb{I}}

\begin{document}


\title{Criteria for system-environment entanglement generation for 
	systems of any size
	in pure-dephasing evolutions}

\author{Katarzyna Roszak}
\affiliation{Department of Theoretical Physics, Faculty of Fundamental Problems of Technology, Wroc{\l}aw University of Science and Technology,
50-370 Wroc{\l}aw, Poland}

\date{\today}

\begin{abstract}
	An evolution between a system and its environment which leads to pure
	dephasing of the system may either be a result of entanglement building
	up between the system and the environment or not (the second option is only possible for initially mixed environmental states). We find
	a way of distinguishing between an entangling and non-entangling evolutions
	for systems which are larger than a single qubit. The generalization of the single qubit
	separability criterion
	to larger systems
	is not sufficient to make this distinction (it constitutes a necessary condition of separability). A set of additional 
	conditions for
	the operators describing the evolution of the environment depending
	on the state of the system is required. We find that the
	commutation of these environmental operators with the initial state
	of the environment does not guarantee separability, products of the operators 
	need to commute among themselves for a pure dephasing evolution
	not to be accompanied by system-environment entanglement generation.
	This is a qualitative difference with respect to the single-qubit case,
	since it allows for a system to entangle with an initially completely mixed
	environment.
\end{abstract}

\maketitle

\section{Introduction}

The detection of system-environment entanglement (SEE) is an involved problem both theoretically
and experimentally once the size of either the system or the environemnt exceeds a couple of qubits \cite{Mintert_PR05,Plenio_QIC07,Horodecki_RMP09,Aolita_RPP15,Kraus_PRA00}. 
In general the quantification of entanglement requires the knowledge of the whole system-environment density matrix.
For open quantum systems with large environments, this is easier to have in theory than in experiment, but is often
also challenging
theoretically, since many of the standard approximate methods are inapplicable in this
case \cite{breuer02,lindblad,maldonado,duffus}, since their aim is to provide the evolution of the system of
interest alone. Only then can the numerical calculation of an entanglement measure be performed
(one that can be found from a density matrix for larger systems, such as
Negativity \cite{vidal02,lee00a,plenio05b}). 
Hence, the studies of entanglement build-up between a system and its 
environment are rather limited in 
literature \cite{eisert02,hilt09,pernice11,pernice12,maziero12,costa16,salamon}
and although they provide insight into the workings of a given open
system, they can be used to draw more general conclusions in a limited
extent.

If the class of systems under study is reduced to such systems for which the interaction with the environment can only lead to pure dephasing of the system of interest, an 
effective theoretical tool to distinguish entangling and non-entangling evolutions has recently become available, as long as the system is only a qubit \cite{roszak15a}. We first extend the results of Ref.~\cite{roszak15a}
and find that qubit-like criteria for system-environment separability
are not sufficient to distinguish between entangling and non-entangling
system-environment evolutions. These criteria 
bind the initial state of the environment to operators which describe
the evolution of the environment conditional on the state of 
the system. We find that, if
any of these criteria is not satisfied
then entanglement is definitely present, but there exist entangled states
for which all of the qubit-like criteria are fulfilled. Furthermore, for
larger systems it is possible for an initially fully mixed environment
to entangle with the qubit during their joint pure-dephasing evolution,
a phenomenon, which is impossible in the case of the qubit.

In fact,
for larger systems
a completely different set of separability criteria is necessary
in addition to the qubit-like conditions, to fully separate
entangled states obtained during the evolution from non-entangled
ones. We find the second set of criteria, which are qualitatively different
from the qubit-like conditions, as they are defined on the conditional
evolution operators of the environment alone.

The paper is organized as follows. In Sec.~\ref{sec1} we describe,
what is meant by ``pure dephasing evolutions'' in terms of the 
system-environment Hamiltonian and write the operators which govern
the joint evolution of the system and environment in this case.
In Sec.~\ref{sec2} we recount the results of Ref.~\cite{roszak15a}
concerning the generation of qubit-environment entanglement, and find
a convenient and general explicitly separable form of any system-environment density 
matrix
which fulfills the separability criterion. Sec.~\ref{sec3} is devoted 
to the study of a qutrit interacting with an environment of any size
and introduces the additional separability criteria for larger systems.
In Sec.~\ref{sec4} a full set of separability criteria for both a system
of any size and an environment of any size is introduced. Sec.~\ref{sec5}
concludes the paper.

\section{Pure dephasing evolutions \label{sec1}}
In the study of the generation of SEE (or to be precise, the distinction
between entangling and non-entangling evolutions) 
for systems which are larger than a qubit,
we remain in the same framework as in the single qubit study \cite{roszak15a}, meaning that we study a class of Hamiltonians
for which the interaction with the environment leads to pure dephasing of the
system of interest (after tracing out the environmental degrees of freedom).
For systems of dimension $N$
and an unspecified, arbitrary size of the environment, 
the class of Hamiltonians 
is
\begin{equation}
\label{ham}
\hat{H}=
\sum_{k=0}^{N-1}\varepsilon_k|k\rangle\langle k|+\hat{H}_{\mathrm{E}}+
\sum_{k=0}^{N-1}|k\rangle\langle k|\otimes{\hat{V}_k},
\end{equation}
where the first term on the right describes the free Hamiltonian of the system
in its eigenbasis $\{|k\rangle\}$, $\hat{H}_{\mathrm{E}}$ is an arbitrary free 
Hamiltonian of the environment, and the last term describes the interaction,
which has to be diagonal in the basis $\{|k\rangle\}$ of the system, while the
environmental operators $\hat{V}_k$ are also arbitrary.

Since the free Hamiltonian of the system commutes with all
other terms of the Hamiltonian and local unitary operations do not 
affect the amount of SEE, in the following
we will always implicitly write the evolution of the system 
(and the corresponding evolution operators) without the unitary oscillations
of the system which result from this first term of the Hamiltonian.
Hence, the density matrix of the system and environment under consideration
is in fact of the form $\hat{U}_S^{\dagger}(t)\hat{\sigma}(t)\hat{U}_S(t)$
with
\begin{equation}
\nonumber
\hat{U}_S(t)=
\exp\left(-\frac{i}{\hbar}t\sum_{k=0}^{N-1}\varepsilon_k|k\rangle\langle k|
\right)
=\sum_{k=0}^{N-1}e^{-\frac{i}{\hbar}\varepsilon_kt}|k\rangle\langle k|.
\end{equation}
This does not affect any of the later conclusions drawn with respect to 
SEE.

The evolution operator for the system and environment
(without the local unitary oscillations of the system) is of the form
\begin{equation}
\label{u}
\hat{U}(t)=\sum_{k=0}^{N-1}|k\rangle\langle k|\otimes{\hat{w}_k(t)},
\end{equation}
with the evolution operator of the environment conditional on the state of the
system given by
\begin{equation}
\label{w}
\hat{w}_k(t)=\exp\left(-\frac{i}{\hbar}(\hat{H}_{\mathrm{E}}+\hat{V}_k)t
\right).
\end{equation}

We assume that the system and environment are initially in a product state, with
the system in a pure state $|\psi\rangle=\sum_{k=0}^{N-1}c_k|k\rangle$, while there are no limitations on the initial
state of the environment, $\hat{R}(0)$. 
Hence the time evolution of the combined system-environment state
is given by
\begin{equation}
\label{sigma}
\hat{\sigma}(t)=\hat{U}(t)\left(|\psi\rangle\langle\psi |\otimes\hat{R}(0)\right)\hat{U}^{\dagger}(t)
=\sum_{k,l=0}^{N-1}c_kc^*_l
|k\rangle\langle l|\hat{R}_{kl}(t),
\end{equation}
where 
\begin{equation}
\label{Rab}
\hat{R}_{kl}(t)=\hat{w}_k(t)\hat{R}(0)\hat{w}_l^{\dagger}(t).
\end{equation}

\section{Separability for a qubit \label{sec2}}

Let us first look at the situation, when our system is a qubit, $N=2$,
following the results of Ref.~\cite{roszak15a}.
In this case the ``if and only if'' condition for separability at time $t$ is 
$[\hat{R}(0),\hat{w}_0^{\dagger}(t)\hat{w}_1(t)]=0$ \cite{roszak15a},
or, equivalently, $\hat{R}_{00}(t)=\hat{R}_{11}(t)$.
Hence, the density matrix of an evolving system and environment
which is separable at time $t$
can always be written as
\begin{equation}
\label{mac1}
\hat{\sigma}(t)=\left(
\begin{array}{cc}
|c_0|^2\hat{R}_{00}(t)&c_0c_1^*
\hat{R}_{01}(t)\\
c_0^*c_1
\hat{R}_{10}(t)&|c_1|^2\hat{R}_{00}(t)
\end{array}\right).
\end{equation}
Here we use a notation, in which the matrix is written in terms of system
states, while the environmental degrees of freedom are contained 
in the operators $\hat{R}_{ij}(t)$, which would be explicitly written
in Dirac notation.
To see that the form of the density matrix given by
eq.~(\ref{mac1}) actually guarantees qubit-environment
separability, it is best to transform the off-diagonal environmental
matrices $\hat{R}_{ij}(t)$, $i\neq j$, in the following way,
\begin{eqnarray}
\nonumber
\hat{R}_{01}(t)&=&\hat{w}_0(t)\hat{R}(0)\hat{w}_0^{\dagger}(t)\hat{w}_0(t)
\hat{w}_1^{\dagger}(t)=\hat{R}_{00}(t)\hat{w}_0(t)
\hat{w}_1^{\dagger}(t),\\
\nonumber
\hat{R}_{10}(t)&=&\hat{w}_1(t)\hat{R}(0)\hat{w}_1^{\dagger}(t)\hat{w}_1(t)
\hat{w}_0^{\dagger}(t)=\hat{R}_{11}(t)\hat{w}_1(t)
\hat{w}_0^{\dagger}(t)\\
\label{rw}
&=&\hat{R}_{00}(t)\hat{w}_1(t)
\hat{w}_0^{\dagger}(t).
\end{eqnarray}

Note, that this procedure can be performed in such a way that the $\hat{w}_k(t)$ operators
are left over on the left side, leading to
\begin{eqnarray}
\nonumber
\hat{R}_{01}(t)&=&\hat{w}_0(t)
\hat{w}_1^{\dagger}(t)\hat{R}_{00}(t),\\
\nonumber
\hat{R}_{10}(t)&=&\hat{w}_1(t)
\hat{w}_0^{\dagger}(t)\hat{R}_{00}(t),
\end{eqnarray}
which shows that the following commutation relations are true
\begin{subequations}
	\label{comm}
\begin{eqnarray}
\left[\hat{R}_{00}(t),\hat{w}_1(t)\hat{w}_0^{\dagger}(t)\right]&=&0,\\
\left[\hat{R}_{00}(t),\hat{w}_0(t)\hat{w}_1^{\dagger}(t)\right]&=&0,\\
\left[\hat{w}_0(t)\hat{w}_1^{\dagger}(t),
\hat{w}_1(t)\hat{w}_0^{\dagger}(t)\right]&=&0.
\end{eqnarray}
\end{subequations}
The last commutation relation comes from the fact that
$\left(\hat{w}_0(t)\hat{w}_1^{\dagger}(t)\right)^{\dagger}=
\hat{w}_1(t)\hat{w}_0^{\dagger}(t)$, which for unitary operators, such as
$\hat{w}_0(t)\hat{w}_1^{\dagger}(t)$ implies commutation (since they can
be diagonalized in the same basis).
Note that there is nothing assumed about the commutation of $\hat{w}_0(t)$ and
$\hat{w}_1(t)$, or $\hat{w}_0(t)\hat{w}_1^{\dagger}(t)$ and
$\hat{w}_0^{\dagger}(t)\hat{w}_1(t)$.

With the use of eqs (\ref{rw}), we can write the whole density matrix
(\ref{mac1}) in the form
\begin{equation}
\label{mac2}
\hat{\sigma}(t)=\left(
\begin{array}{cc}
|c_0|^2\hat{R}_{00}(t)&
c_0c_1^*\hat{R}_{00}(t)\hat{w}_0(t)\hat{w}_1^{\dagger}(t)\\
c_0^*c_1\hat{w}_1(t)\hat{w}_0^{\dagger}(t)\hat{R}_{00}(t)&
|c_1|^2\hat{R}_{00}(t)
\end{array}\right).
\end{equation}
Since the commutation relations (\ref{comm}) imply that there exists
an environmental basis $\{|n(t)\rangle\}$ which diagonalizes
$\hat{R}_{00}(t)$, $\hat{w}_0(t)\hat{w}_1^{\dagger}(t)$, and
$\hat{w}_1(t)\hat{w}_0^{\dagger}(t)$ at time $t$, we can write all three operators
with the help of this basis,
\begin{subequations}
\begin{eqnarray}
\hat{R}_{00}(t)&=&\sum_np_n(t)|n(t)\rangle\langle n(t)|,\\
\hat{w}_0(t)\hat{w}_1^{\dagger}(t)&=&\sum_n e^{i\phi_n(t)}|n(t)\rangle\langle n(t)|,\\
\hat{w}_1(t)\hat{w}_0^{\dagger}(t)&=&\sum_n e^{-i\phi_n(t)}|n(t)\rangle\langle n(t)|.
\end{eqnarray}
\end{subequations}
Now the density matrix (\ref{mac2})
can be written with the help of basis $\{|n(t)\rangle\}$
in an obviously separable form,
\begin{equation}
\label{sigmasep}
\hat{\sigma}(t)=\sum_n p_n(t)\left(
\begin{array}{cc}
|c_0|^2&c_0c_1^*e^{i\phi_n(t)}\\
c_0^*c_1e^{-i\phi_n(t)}&|c_1|^2
\end{array}
\right)\otimes |n(t)\rangle\langle n(t)|.
\end{equation}
With the density matrix in this form, it is also obvious that it is 
zero-discordant with respect to the environment, but not necessarily
with respect to the qubit \cite{roszak17}.

\section{Separability of a qutrit \label{sec3}}
\subsection{The inadequacy of the qubit-like 
	separability conditions alone \label{sec3a}}

The previous section allows us to easily find the only possible 
separable form of 
the qubit-environment density matrix for pure dephasing evolutions.
Furthermore, it will allow us to extend the reasoning to systems
of dimension $N>2$,
and check, if a straightforward extension of the qubit separability
condition to bigger systems is enough to guarantee separability for
$N>2$.

Let us for now restrict ourselves to an $N=3$ system (a qutrit) interacting with
the environment as described in Sec.~\ref{sec1}. The idea is not
to repeat the reasoning of Ref.~\cite{roszak15a} and look at minors; instead, we will look at the time-evolved qutrit-environment
density matrix (\ref{sigma}),
\begin{equation}
\label{sigma3}
\hat{\sigma}(t)=\left(\begin{array}{ccc}
|c_0|^2\hat{R}_{00}(t)&c_0c_1^*\hat{R}_{01}(t)&c_0c_2^*\hat{R}_{02}(t)\\
c_1c_0^*\hat{R}_{10}(t)&|c_1|^2\hat{R}_{11}(t)&c_1c_2^*\hat{R}_{12}(t)\\
c_2c_0^*\hat{R}_{20}(t)&c_2c_1^*\hat{R}_{21}(t)&|c_2|^2\hat{R}_{22}(t)\\
\end{array}
\right),
\end{equation}
and impose the conditions, which are an extension of the 
separability conditions for a qubit,
\begin{equation}
\label{war3}
[\hat{R}(0),\hat{w}_0^{\dagger}\hat{w}_1]=
[\hat{R}(0),\hat{w}_1^{\dagger}\hat{w}_2]=
[\hat{R}(0),\hat{w}_2^{\dagger}\hat{w}_0]=0,
\end{equation}
or, equivalently, 
$\hat{R}_{00}(t)=\hat{R}_{11}(t)=\hat{R}_{22}(t)$.
Under these conditions, we get (as in eq.~(\ref{rw}))
\begin{equation}
\hat{R}_{ij}(t)=\hat{R}_{00}(t)\hat{w}_i(t)\hat{w}_j^{\dagger}(t)
\end{equation}
and
\begin{subequations}
\begin{eqnarray}
\label{15a}
\left[\hat{R}_{00}(t),\hat{w}_i(t)\hat{w}_j^{\dagger}(t)\right]&=&0,\\
\left[\hat{R}_{00}(t),\hat{w}_j(t)\hat{w}_i^{\dagger}(t)\right]&=&0,\\
\label{ostatnia}
\left[\hat{w}_j(t)\hat{w}_i^{\dagger}(t),\hat{w}_i(t)\hat{w}_j^{\dagger}(t)\right]&=&0,
\end{eqnarray}
\end{subequations}
for all $i$ and $j$.

Hence, if the density matrix of a qutrit and its environment 
(\ref{sigma3}) fulfills the conditions (\ref{war3}), it
can be written as
\begin{widetext}
\begin{equation}
\label{sigma4}
\hat{\sigma}(t)=\left(\begin{array}{ccc}
|c_0|^2\hat{R}_{00}(t)
&c_0c_1^*\hat{R}_{00}(t)\hat{w}_0(t)\hat{w}_1^{\dagger}(t)
&c_0c_2^*\hat{R}_{00}(t)\hat{w}_0(t)\hat{w}_2^{\dagger}(t)\\
c_1c_0^*\hat{w}_1(t)\hat{w}_0^{\dagger}(t)\hat{R}_{00}(t)
&|c_1|^2\hat{R}_{00}(t)
&c_1c_2^*\hat{R}_{00}(t)\hat{w}_1(t)\hat{w}_2^{\dagger}(t)\\
c_2c_0^*\hat{w}_2(t)\hat{w}_0^{\dagger}(t)\hat{R}_{00}(t)
&c_2c_1^*\hat{w}_2(t)\hat{w}_1^{\dagger}(t)\hat{R}_{00}(t)
&|c_2|^2\hat{R}_{00}(t)\\
\end{array}
\right).
\end{equation}
\end{widetext}
This density matrix is not necessarily always separable, because although for all
$i\neq j$ there exists a basis $\{|n_{ij}(t)\rangle\}$ which diagonalizes
$\hat{R}_{00}(t)$, $\hat{w}_j(t)\hat{w}_i^{\dagger}(t)$, and $\hat{w}_i(t)\hat{w}_j^{\dagger}(t)$, there is no reason for the different bases
$\{|n_{ij}(t)\rangle\}$ to be the same. Hence,
unless we also have
\begin{equation}
	\label{wardod}
\left[\hat{w}_i(t)\hat{w}_j^{\dagger}(t),
\hat{w}_k(t)\hat{w}_l^{\dagger}(t)\right]=0
\end{equation}
for all $i,j,k,l=0,1,2$, it is impossible to write the density matrix (\ref{sigma4}) in the simple separable
form as done in eq.~(\ref{sigmasep}) at the end of Sec.~\ref{sec2}.
The conditions (\ref{wardod}) actually come down to one independent,
 nontrivial equation; it should be obvious that there are three such
 equations taking into account the commutation relation (\ref{ostatnia}),
 since there are three relevant combinations of indexes $01$, $02$, and
 $12$,
 while the reduction to one will become apparent in Sec.~\ref{sec3c}.

It is fairly simple to show that, if any of the conditions of eq.~(\ref{war3}) are not fulfilled, then the system is entangled
with its environment following the method used in Ref.~\cite{roszak15a}.
The proof of this fact
for a system of any size is provided in Appendix \ref{appA}.

Hence, if condition (\ref{war3}) is broken, it indicates entanglement,
while if both condition (\ref{war3}) and conditions (\ref{wardod}) are 
fulfilled, the system and environment are separable. It is the gray area
where condition (\ref{war3}) is fulfilled, while any of the conditions (\ref{wardod})
are broken, that is interesting. There are three possibilities:
\begin{itemize}
	\item The difference between the qubit and qutrit situations is only apparent, and the fulfillment of condition (\ref{war3}) is both
	necessary and sufficient for separability.
	\item The fulfillment of condition (\ref{war3}) is 
	necessary for separability, while the conditions (\ref{wardod}) are necessary and sufficient for separability (so when any of the conditions (\ref{wardod})
	are not met, there is entanglement in the system).
	\item The fulfillment of condition (\ref{war3}) is 
	necessary for separability, while the conditions (\ref{wardod}) are sufficient (so when the conditions (\ref{wardod})
	are met, there is no entanglement in the system, but there are also 
	situations, when the conditions (\ref{wardod}) are not fulfilled
	and there is no SEE).
\end{itemize} 
	The following example will eliminate the first option,
	since it will show that there exists a situation when a system is entangled with its
	environment, even though condition (\ref{war3}) is fulfilled.

\subsection{Example: Entanglement of a qutrit with a initial completely mixed state of the environment \label{sec3b}}

Let us look at an exemplary situation, in which condition (\ref{war3})
is fulfilled, but some of the conditions (\ref{wardod}) are not.
To make the example as simple as possible we will consider a
single-qubit environment (the system is a qutrit). We assume that the
initial state of the qutrit is an equal superposition state
$|\psi\rangle=1/\sqrt{3}(|0\rangle +|1\rangle +|2\rangle)$ for simplicity.
The initial state of the environment is a completely mixed state
$\hat{R}(0)=1/2(|0\rangle\langle 0|+|1\rangle\langle 1|)$, which guarantees
the fulfillment condition (\ref{war3}), since the unit matrix commutes
with all other matrices.

We are interested in a certain instant of time $t$, 
when the conditional evolution operators of the environment
are proportional to unity and two of the Pauli matrices,
\begin{subequations}
\begin{eqnarray}
\hat{w}_0(t)&=&\mathds{1},\\
\hat{w}_1(t)&=&
\left(\begin{array}{cc}
1&0\\
0&-1
\end{array}
\right),\\
\hat{w}_2(t)&=&
\left(\begin{array}{cc}
0&i\\
-i&0
\end{array}
\right),
\end{eqnarray}
\end{subequations}
so $\hat{w}_0(t)$ commutes with $\hat{w}_1(t)$ and $\hat{w}_2(t)$
(all three are hermitian),
but $\hat{w}_1(t)$ and $\hat{w}_2(t)$ do not commute with each other,
nor do they commute with
\begin{equation}
\hat{w}_1(t)\hat{w}_2(t)=
\left(\begin{array}{cc}
0&i\\
i&0
\end{array}
\right),
\end{equation}
which is proportional to the thrid Pauli matrix.
 Hence the system-environment density matrix at time $t$ is given by
 \begin{subequations}
 \begin{eqnarray}
 \hat{\sigma}(t)&=&\frac{1}{6}
 \left(\begin{array}{ccc}
 \mathds{1}&\hat{w}_1^{\dagger}(t)&\hat{w}_2^{\dagger}(t)\\
 \hat{w}_1(t)&\mathds{1}&\hat{w}_1\hat{w}_2^{\dagger}(t)\\
 \hat{w}_2(t)&\hat{w}_2(t)\hat{w}_1^{\dagger}(t)&\mathds{1}
 \end{array}
 \right)\\
 \label{mac}
 &=&\frac{1}{6}
 \left(\begin{array}{cccccc}
 1&0&1&0&0&i\\
 0&1&0&-1&-i&0\\
 1&0&1&0&0&i\\
 0&-1&0&1&i&0\\
 0&i&0&-i&1&0\\
 -i&0&-i&0&0&1\\
 \end{array}
 \right).
 \end{eqnarray}
 \end{subequations}
 The matrix in eq.~(\ref{mac}) is written explicitly in the system-environment basis in the following order
 $\{|00\rangle, |01\rangle, |10\rangle, |11\rangle, |20\rangle, |21\rangle\}$,
 where in the notation $|se\rangle=|s\rangle\otimes |e\rangle$,
 $s$ denotes the state of the system, while $e$ denotes the state
 of the environment.
 
 To check, if there is SEE in the state described
 by the density matrix (\ref{mac}), it suffices to find the eigenvalues
 of the matrix after partial transposition with respect to either the 
 system or the environment. If the transposed matrix is not a density 
 matrix (has negative eigenvalues) then there is entanglement in the system
 (as stated by the Peres-Horodecki criterion \cite{Peres_PRL96,Horodecki_PLA96}). 
 After partial transposition with respect to the environment 
 we get 
 \begin{equation}
 \label{macT}
 \hat{\sigma}^{T_E}(t)=\frac{1}{6}
 \left(\begin{array}{cccccc}
 1&0&1&0&0&-i\\
 0&1&0&-1&i&0\\
 1&0&1&0&0&i\\
 0&-1&0&1&i&0\\
 0&-i&0&-i&1&0\\
 i&0&-i&0&0&1\\
 \end{array}
 \right).
 \end{equation}
 The matrix has two negative eigenvalues equal to $(-\frac{1}{6})$ 
 (and four positive eigenvalues equal to $\frac{1}{3}$), so it is not
 a density matrix, and there is entanglement between the system and 
 the environment in the (not transposed) density matrix (\ref{mac}).
 
 There are a couple of interesting conclusions, which can be drawn
 from the above example. The first obviously is that the fulfillment of
 condition (\ref{war3}) is not sufficient for separability. The second,
 which should be surprising, is that pure-dephasing evolution can lead
 to entanglement with the enviroment even, if the environment is initially
 in a completely mixed state (hence, fully classical). This is absolutely not possible, if the system
 is a qubit.
 
 \subsection{The commutation conditions between different
 	environmental evolution operators are necessary and sufficient
 	for separability \label{sec3c}}
 
 In the following we will show that systems for which the qubit-like
 condition (\ref{war3}) is met, but at least one of the conditions
 (\ref{wardod}) is not, display SEE.
 To this end, we study the qutrit-environment density matrix (\ref{sigma4}),
 which is already written in a form equivalent to the fulfillment
 of the qubit-like conditions (\ref{war3}).
 
 For simplicity, let us denote
 \begin{equation}
 \label{notacjaW}
 \hat{W}_{ij}(t)=\hat{w}_i(t)\hat{w}_j^{\dagger}(t)=
 \hat{W}_{ji}^{\dagger}(t),
 \end{equation}
 which automatically implies 
 $\hat{W}_{12}(t)=\hat{W}_{10}(t)\hat{W}_{20}^{\dagger}(t)$.
 It is now convenient to write $\hat{R}_{00}(t)$, $\hat{W}_{10}(t)$ and $\hat{W}_{20}(t)$ in the common eigenbasis of 
 $\hat{R}_{00}(t)$ and $\hat{W}_{10}(t)$, $\{|n_{01}(t)\rangle\}$
 (this basis may change with time),
 \begin{subequations}
 \begin{eqnarray}
 \hat{R}_{00}(t)&=&\sum_n p_n|n_{01}(t)\rangle\langle n_{01}(t)|,\\
 \hat{W}_{10}(t)&=&\sum_n e^{i\phi_n(t)}|n_{01}(t)\rangle\langle n_{01}(t)|,\\
 \hat{W}_{20}(t)&=&\sum_{nm} x_{nm}|n_{01}(t)\rangle\langle m_{01}(t)|.
 \end{eqnarray}
 \end{subequations}
 Obviously $\hat{W}_{20}(t)$ is not diagonal in this basis. 
 If that were the case
 then also $\hat{W}_{10}(t)\hat{W}_{20}^{\dagger}(t)$ would be diagonal in the
 same basis and the state would have to be separable, since
 the conditions (\ref{wardod}) would be fulfilled.
 Furthermore, it should now be evident that, if one of the conditions
 (\ref{wardod}) is met, it implies the fulfillment of the other two,
 since from
 \begin{equation}
 \label{duzew}
 \left[\hat{w}_1(t)\hat{w}_0^{\dagger}(t),\hat{w}_0(t)\hat{w}_2^{\dagger}(t)\right]= \left[\hat{W}_{10}(t),\hat{W}_{20}^{\dagger}(t)\right]=0,
 \end{equation}
 we get
 \begin{eqnarray}
 \nonumber
 \left[\hat{w}_0(t)\hat{w}_1^{\dagger}(t),\hat{w}_1(t)\hat{w}_2^{\dagger}(t)\right]&=& \left[\hat{W}_{10}^{\dagger}(t),\hat{W}_{10}(t)\hat{W}_{20}^{\dagger}(t)\right]=0,\\
 \nonumber
 \left[\hat{w}_2(t)\hat{w}_0^{\dagger}(t),\hat{w}_1(t)\hat{w}_2^{\dagger}(t)\right]&=& \left[\hat{W}_{20}(t),\hat{W}_{10}(t)\hat{W}_{20}^{\dagger}(t)\right]=0.
 \end{eqnarray}
 
 It is now possible to write the whole density matrix (\ref{sigma4})
 at time $t$ using the environmental basis $\{|n_{01}(t)\rangle\}$, which
 will be denoted as simply $\{|n(t)\rangle\}$ in what follows.
 This yields
 \begin{widetext}
 	\begin{equation}
 	\label{sigma5}
 	\hat{\sigma}(t)=\left(\begin{array}{ccc}
 	|c_0|^2\sum_n p_n|n(t)\rangle\langle n(t)|
 	&c_0c_1^*\sum_n p_ne^{-i\phi_n(t)}|n(t)\rangle\langle n(t)|
 	&c_0c_2^*\sum_{nm} p_nx_{mn}^*|n(t)\rangle\langle m(t)|\\
 	c_1c_0^*\sum_n p_ne^{i\phi_n(t)}|n(t)\rangle\langle n(t)|
 	&|c_1|^2\sum_n p_n|n(t)\rangle\langle n(t)|
 	&c_1c_2^*\sum_{nm} p_ne^{i\phi_n(t)}x_{mn}^*|n(t)\rangle\langle m(t)|\\
 	c_2c_0^*\sum_{nm} p_mx_{nm}|n(t)\rangle\langle m(t)|
 	&c_2c_1^*\sum_{nm} p_me^{-i\phi_m(t)}x_{nm}|n(t)\rangle\langle m(t)|
 	&|c_2|^2\sum_n p_n|n(t)\rangle\langle n(t)|\\
 	\end{array}
 	\right).
 	\end{equation}
 \end{widetext}
 Note that in this basis, five out of nine submatrices
 describing the environment conditional on the element of the density
 matrix of the system
 are diagonal.
 
 To check, if the density matrix is entangled, we again use the 
 Peres-Horodecki criterion \cite{Peres_PRL96,Horodecki_PLA96}, firstly applying a partial transposition
 to eq.~(\ref{sigma5}) with respect to the system and then checking,
 if the resulting matrix has negative eigenvalues
 (if it does, then there is entanglement between the system and the 
 environment, otherwise the question is unanswered). 
 The partial transposition
 yields
 \begin{widetext}
 	\begin{equation}
 	\label{transpose}
 	\hat{\sigma}^{T_S}(t)=\left(\begin{array}{ccc}
 	|c_0|^2\sum_n p_n|n(t)\rangle\langle n(t)|
 	&c_0c_1^*\sum_n p_ne^{i\phi_n(t)}|n(t)\rangle\langle n(t)|
 	&c_2c_0^*\sum_{nm} p_mx_{nm}|n(t)\rangle\langle m(t)|\\
 	c_1c_0^*\sum_n p_ne^{-i\phi_n(t)}|n(t)\rangle\langle n(t)|
 	&|c_1|^2\sum_n p_n|n(t)\rangle\langle n(t)|
 	&c_2c_1^*\sum_{nm} p_me^{-i\phi_m(t)}x_{nm}|n(t)\rangle\langle m(t)|\\
 	c_0c_2^*\sum_{nm} p_nx_{mn}^*|n(t)\rangle\langle m(t)|
 	&c_1c_2^*\sum_{nm} p_ne^{i\phi_n(t)}x_{mn}^*|n(t)\rangle\langle m(t)|
 	&|c_2|^2\sum_n p_n|n(t)\rangle\langle n(t)|\\
 	\end{array}
 	\right).
 	\end{equation}
 \end{widetext}
 Since the diagonalization of the matrix (\ref{transpose}) for an 
 arbitrary size of the environment is impossible, we restrict ourselves
 to the study of its principal minors, since a matrix has only non-negative
 eigenvalues, if and only if all of its principal minors are non-negative.
 
 We have identified a class of principal minors, which solve the
 question of qubit-environment entanglement generation, when the conditions
 (\ref{wardod}) are not fulfilled. These minors are determinants of 
 $3\times 3$ matrices obtained by symmetrically crossing out $(3M-3)$
 rows and columns out of the transposed matrix (\ref{transpose}),
 where $M$ is the dimension of the environment. They are described by 
 two indices $k,q=0,1,...,M-1$, which correspond to two states of 
 the environment.
 Note, that the rows and columns of the matrix (\ref{transpose})
 which is of dimension $3M\times 3M$
 are numbered by both the system and the environmental states
 and are ordered in such a way, that the row/column number
 $r=sM+e$, where $s=0,1,2$ denote system states, while $e=0,1,...,M-1$
 denote states of the environment. To obtain the $3\times 3$ matrix for a given $k$ and $q$,
 we cross out all rows and columns with the exception of the $k$-th,
 $(M+k)$-th and $(2M+q)$-th rows and columns. This means that only the
 $k$-th diagonal elements from the parts of the matrix (\ref{transpose}) 
 proportional to $|c_0|^2$ and $|c_1|^2$ are left over, as well as
 the $q$-th element from the part proportional to $|c_1|^2$, which yields
 principal minors of the form
 \begin{widetext}
 \begin{equation}
 D_{kq}=\det\left(
 \begin{array}{ccc}
 |c_0|^2p_k&c_0c_1^*p_ke^{i\phi_k(t)}&c_2c_0^*p_qx_{kq}\\
 c_0^*c_1p_ke^{-i\phi_k(t)}&|c_1|^2p_k&c_2c_1^*p_qe^{-i\phi_q(t)}x_{kq}\\
 c_2^*c_0p_qx_{kq}^*&c_2^*c_1p_qe^{i\phi_q(t)}x_{kq}^*&|c_2|^2p_q
 \end{array}
 \right).
 \end{equation}
 \end{widetext}
 
 Since from the condition (\ref{war3}) we get eq.~(\ref{15a})
 and consequently
 $\left[\hat{R}_{00},\hat{W}_{20}\right]=0$, this means
 that either $p_k=p_q$ or $x_{kq}=0$. If $x_{kq}=0$, then
 the principal minor $D_{kq}=0$.
 Otherwise we get 
 \begin{equation}
 D_{kq}=-2|c_0c_1c_2|^2p_k^3|x_{kq}|^2\left[1-\cos\left(
 \phi_k(t)-\phi_q(t)\right)\right],
 \end{equation}
 which cannot be positive, and is equal to zero only
 in four situations . Two of them are trivial. The first is when
 the initial occupation of the qubit $|c_i|^2=0$ for any $i=0,1,2$,
 which means that the studied system is operationally a qubit, not a 
 qutrit. In the second trivial situation both of the eigenvalues of 
 the matrix $\hat{R}_{00}(t)$ are zero,
 $p_k=p_q=0$, so regardless of the value of the parameter $x_{kq}$,
 there are no transitions between the states $|k(t)\rangle $ and $|q(t)\rangle$.
 In this situation, no transitions will occur to states 
 $|k(t)\rangle $ or $|q(t)\rangle$, since for any state $|n(t)\rangle$,
 either $p_n=p_{k/q}$ and must be equal to zero, or $x_{(k/q)n}=0$.
 Hence, the density matrix elements corresponding to the states
 $|k(t)\rangle $ and $|q(t)\rangle$ must be equal to zero during any point 
 of the evolution, and $|k(t)\rangle $ and $|q(t)\rangle$ can be eliminated
 from the subspace of environmental degrees of freedom taken into account.
 Note that the discussed system-environment state is already restricted
 by the fulfillment of conditions (\ref{war3}); the above conclusion
 is by no means general.
 
 The other two situations are relevant, if we wish to distinguish between
 entangling and non-entangling evolutions. For the determinant $D_{kq}$
 to be zero, we must either have $x_{kq}=0$ or
 $e^{i\phi_k(t)}=e^{i\phi_q(t)}=0$, otherwise $D_{kq}<0$ which means
 that there is entanglement in the system. If it is true for all $k$ and $q$
 that $x_{kq}=0$ or
 $e^{i\phi_k(t)}-e^{i\phi_q(t)}=0$, then it is equivalent to write
 \begin{equation}
 \label{wardod2}
 \left[\hat{W}_{10}(t),\hat{W}_{20}(t)\right]=0,
 \end{equation}
 which means that eq.~(\ref{duzew}) is also fulfilled, and consequently
 the conditions (\ref{wardod}) are met. In this case we can show, using
 the definition of separability that the qutrit is separable from
 its environment. Otherwise there is qutrit-environment entanglement
 in the studied state.
 
 Hence, we have shown that both the condition (\ref{war3}) and the
 condition (\ref{wardod2}) must be fulfilled for a qutrit to be separable
 from its environment. If either of the conditions are not met, then
 there is entanglement in the system. Note that for a qutrit coupled
 to a qubit environment (as in the example of Sec.~\ref{sec3b}), the
 ``grey area'' where the condition (\ref{war3}) is met, but condition (\ref{wardod2}) is not, contains only situations with an initially 
 fully classical environment $p_0=p_1=1/2$, because $p_0\neq p_1$
 implies $x_{01}=x_{10}=0$, for which two-dimensional operators 
 $\hat{W}_{10}(t)$ and $\hat{W}_{20}(t)$
 must commute. For larger environments, other, more quantum initial
 states of the environment may lead to the generation of this kind of 
 entanglement.
 
 \section{Larger systems \label{sec4}}
 There is no qualitative difference when studying systems of larger
 dimensionality, $N>3$, compared to the qutrit case, as there is
 between a qubit and a qutrit. The qubit-like conditions, correspondnig
 to those given by eq.~(\ref{war3}) for the qutrit, now take the form
 that for all $i,j=0,1,...,N-1$ (the indices label system states)
 \begin{equation}
 \label{warN}
 [\hat{R},\hat{w}_i^{\dagger}(t)\hat{w}_j(t)]=0.
 \end{equation}
 Obviously, if the condition
 (\ref{warN}) is satisfied for $i$ and $j$, then it must
 be satisfied for $j$ and $i$ (interchanged indices)
 and the conditions are automatically fulfilled for $i=j$.
 Furthermore, since for a given $i$ and $j$ the condition of eq.~(\ref{warN})
 is equivalent
 to $\hat{R}_{ii}(t)=\hat{R}_{jj}(t)$, it is evident that if condition
 (\ref{warN})
 is satisfied for $i$ and $j$, as well as for $i$ and $k$, then
 it must be satisfied for $j$ and $k$ as well.
 Hence, there are $(N-1)$ non-trivial qubit-like commutation conditions
 in eq.~(\ref{warN}). If any of these conditions is broken, then there is
 entanglement between the system and its environment in the density 
 matrix given by eq.~(\ref{sigma}). A proof of this fact is given
 in Appendix (\ref{appA}).
 
 When all of the conditions (\ref{warN}) are satisfied, the question of
 separability is still open and another set of conditions needs to be 
 verified. These are similar to the conditions introduced for the qutrit
 scenario (\ref{wardod}), namely that for all $i$, $j$, $k$, and $l$
 (where the indices again label system states only)
 \begin{equation}
 \label{wardodN}
 \left[\hat{W}_{ij}(t),\hat{W}_{kl}(t)\right]=0,
 \end{equation}
 where the environmental operators $\hat{W}_{ij}(t)$ are given by
 eq.~(\ref{notacjaW}).
 
 Note that only $(N-1)(N-2)/2$ of these conditions are
 independent. This is obvious, if we first set two of the system indices
 to a fixed an equal value, say $j=k=0$. It is then straightforward to show
 that if for all $i\neq l$, the conditions
 \begin{equation}
 \label{warN0}
 \left[\hat{W}_{i0}(t),\hat{W}_{0l}(t)\right]=0,
 \end{equation}
 are fulfilled, also all conditions (\ref{wardodN}) must be satisfied.
 To this end, we can write
 \begin{eqnarray}
 \nonumber
 \left[\hat{W}_{ij}(t),\hat{W}_{kl}(t)\right]&=&
 \hat{W}_{ij}(t)\hat{W}_{kl}(t)-\hat{W}_{ij}(t)\hat{W}_{kl}(t)\\
 \nonumber&=&
 \hat{W}_{i0}\hat{W}_{0j}\hat{W}_{k0}(t)\hat{W}_{0l}(t)-\hat{W}_{ij}(t)\hat{W}_{kl}(t)\\
 \nonumber&=&
 \hat{W}_{k0}\hat{W}_{0l}\hat{W}_{i0}(t)\hat{W}_{0j}(t)-\hat{W}_{ij}(t)\hat{W}_{kl}(t)\\
 \nonumber&=&\hat{W}_{ij}(t)\hat{W}_{kl}(t)-\hat{W}_{ij}(t)\hat{W}_{kl}(t)
 =0,
 \end{eqnarray}
 since $\hat{W}_{i0}(t)=\hat{W}_{0i}^{\dagger}(t)$ and the operators 
 are unitary (so the commutation relations of eq.~(\ref{warN0}) hold, if one or both operators undergo hermitian conjugation).
 
 \section{Conclusion and outlook \label{sec5}}
 We have shown that there is a qualitative difference between 
 system-environment generation in case of larger systems and in case
 of a qubit. Although a generalization of the qubit separability
 condition to larger systems does constitute an entanglement witness
 (if any of the qubit-like conditions is broken, this means that
 there is entanglement in the system), entanglement can be generated
 also, if all of such conditions are met. 
 
 Furthermore, we have identified 
 an additional set of separability conditions which apply for systems
 of larger dimensionality than a qubit, when all qubit-like conditions
 are fulfilled. These conditions act on evolution
 operators which govern environment behavior depending on different states
 of the system, and contrarily to the qubit-like conditions, are 
 independent of the actual initial state of the environment
 (neither set of conditions is dependent on the system state, as long as
 it is a superposition of pointer states). This set of
 conditions allows us to make a final distinction between entangling
 and non-entangling evolutions, since if they are all fulfilled
 at a given time
 (additionally to the qubit-like conditions), the system-environment 
 state at this time is separable. Otherwise it is always entangled.
 
 The number of separability conditions expectedly grows with the size of
 the studied system. There are $N-1$ conditions of the first type and
 $(N-1)(N-2)/2$ conditions of the second type for a system of size $N$.
 Hence, the number of qubit-like conditions grows more slowly, but the number
 of conditions of the other type is smaller for a both a qubit and a qutrit,
 while for $N=4$ there are three non-trivial conditions of each type.
 Still, for reasonably small systems, and/or interactions of large symmetry,
 checking entanglement generation using the proposed method should
 be manageable.
 
 \section*{Acknowledgements}
 K.~R.~would like to thank Dariusz Chru{\'s}ci{\'n}ski and {\L}ukasz
 Cywi{\'n}ski for useful discussions.

 \begin{appendix}
 	\section{Proof of entanglement generation when any of the qubit-like
 		conditions is broken \label{appA}}
 	
 	Let us consider a system of any size $N$ interacting with an environment
 	of any size $M$ via a pure-dephasing interaction described in
 	Sec.~\ref{sec1}. The time-evolved density matrix of the system
 	and its environment can be written using eq.~(\ref{sigma}) and
 	after partial transposition with respect to the system, we get
 	\begin{equation}
 	\label{transposeN}
 	\hat{\sigma}^{T_S}(t)
 	=\sum_{k,l=0}^{N-1}c_kc^*_l
 	|l\rangle\langle k|\hat{R}_{kl}(t),
 	\end{equation}
 	where the states $|l\rangle$ and $|k\rangle$ are interchanged
 	with respect to (\ref{sigma}).
 	
 	To prove that there is entanglement between the system and its environment in all situations when at least
 	one of the qubit like conditions given by eq.~(\ref{warN})
 	is broken, it is convenient to write the conditions in an equivalent form,
 	\begin{equation}
 	\label{warNequiv}
 	\left[\hat{R}_{ii}(t),\hat{W}_{ij}(t)\right]=0,
 	\end{equation}
 	where 
 	$\hat{R}_{ii}(t)$ is given by eq.~(\ref{Rab}) and $\hat{W}_{ij}(t)$
 	is given by eq.~(\ref{notacjaW}).
 	
 	To prove the equivalence of conditions (\ref{warN}) and (\ref{warNequiv}),
 	we first provide the derivation of (\ref{warNequiv}) from (\ref{warN}).
 	Since we may always write
 	$\hat{R}_{ij}(t)=\hat{R}_{ii}(t)\hat{W}_{ij}(t)=
 	\hat{W}_{ij}(t)\hat{R}_{jj}(t)$
 	and the condition (\ref{warN}) can easily be transformed into
 	$\hat{R}_{ii}(t)=\hat{R}_{jj}(t)$, we get 
 	$\hat{R}_{ii}(t)\hat{W}_{ij}(t)=
 	\hat{W}_{ij}(t)\hat{R}_{ii}(t)$.
 	For the derivation of (\ref{warN}) from (\ref{warNequiv}), we start
 	by writing eq.~(\ref{warNequiv}) explicitly in terms of
 	the $\hat{w}_i(t)$ operators and the initial state of the environment
 	$\hat{R}(0)$,
 	\begin{equation}
 	\hat{w}_i(t)\hat{R}(0)\hat{w}^{\dagger}_i(t)\hat{w}_i(t)\hat{w}^{\dagger}_j(t)-\hat{w}_i(t)\hat{w}^{\dagger}_j(t)\hat{w}_i(t)\hat{R}(0)\hat{w}^{\dagger}_i(t)=0.
 	\end{equation}
 	Since $\hat{w}_i(t)$ are unitary, multiplying on the right by $\hat{w}_i(t)$ yields
 	\begin{equation}
 	\hat{w}_i(t)\left(\hat{R}(0)\hat{w}^{\dagger}_j(t)\hat{w}_i(t)-
 	\hat{w}^{\dagger}_j(t)\hat{w}_i(t)\hat{R}(0)\right)=0,
 	\end{equation}
 	and multiplying by $\hat{w}^{\dagger}_i(t)$ on the left yields
 	the condition (\ref{warN}). Hence, showing that the violation of any
 	of the conditions (\ref{warNequiv}) guarantees system-environment 
 	entanglement is equivalent to showing the same for the violation
 	of any of the conditions (\ref{warN}).
 	
 	To show that, if for any $i\neq j$ there is $\left[\hat{R}_{ii}(t),\hat{W}_{ij}(t)\right]\neq 0$, then
 	there is system-enviroment entanglement at time $t$, we use the
 	Peres-Horodecki criterion \cite{Peres_PRL96,Horodecki_PLA96}. Hence, we must show that in this case
 	the system-environment density matrix
 	after partial transposition (\ref{transposeN}) has negative eigenvalues.
 	Since a matrix has negative eigenvalues, if any of its principal minors
 	is negative, we study a class of minors which is indicative
 	of this type of entanglement, extending the results of Ref.~\cite{roszak15a}. For given system states $i$ and $j$ we first 
 	eliminate all elements of the matrix (\ref{transposeN}) describing
 	other system states, by symmetrically crossing out rows and columns
 	denoted by other system indices than $ii$, $ij$, $ji$, and $jj$.
 	The resulting matrix is of the form
 	\begin{equation}
 	\label{Mij}
 	\hat{M}_{ij}=\left(
 	\begin{array}{cc}
 	|c_i|^2\hat{R}_{ii}(t)&c_i^*c_j\hat{W}_{ij}^{\dagger}(t)\hat{R}_{ii}(t)\\
 	c_ic_j^2\hat{R}_{ii}(t)\hat{W}_{ij}(t)
 	&|c_j|^2\hat{W}_{ij}^{\dagger}(t)\hat{R}_{ii}(t)\hat{W}_{ij}(t)
 	\end{array}
 	\right),
 	\end{equation}
 	since
 	\begin{subequations}
 		\begin{eqnarray}
 		\hat{R}_{jj}(t)&=&\hat{W}_{ij}^{\dagger}(t)\hat{R}_{ii}(t)\hat{W}_{ij}(t),\\
 		\hat{R}_{ij}(t)&=&\hat{R}_{ii}(t)\hat{W}_{ij}(t),\\
 		\hat{R}_{ji}(t)&=&\hat{W}_{ij}^{\dagger}(t)\hat{R}_{ii}(t).
 		\end{eqnarray}
 	\end{subequations}
 	
 	We will not study the minor corresponding to the matrix (\ref{Mij}),
 	since
 	for an arbitrary size of the environment, calculating it is too
 	complex. Instead we write the matrix in the eigenbasis of  
 	$\hat{R}_{ii}(t)$, which is specific for time $t$, and which we
 	denote as $\{|n(t)\rangle\}$. In this basis, we have 
 	\begin{subequations}
 		\begin{eqnarray}
 		\hat{R}_{ii}(t)&=&\sum_np_n(t)|n(t)\rangle\langle n(t)|,\\
 		\hat{W}_{ij}(t)&=&\sum_{nm}y_{nm}(t)|n(t)\rangle\langle m(t)|,
 		\end{eqnarray}
 	\end{subequations}
 	where $p_n(t)$ are the eigenvalues of $\hat{R}_{ii}(t)$ and
 	$y_{nm}(t)=\langle n(t)|\hat{W}_{ij}(t)|m(t)\rangle$.
 	The indices $i$ and $j$ are omitted here.
 	Now the class of relevant principal minors is obtained by 
 	symmetrically crossing out $M-1$ rows and columns from the
 	matrix (\ref{Mij}) in such a way, that only one diagonal element
 	proportional to $|c_{jj}|^2$ is left (and then finding the determinant).
 	Hence, we get $M$ principal minors in this class labeled by 
 	the environmental state index $n=0,1,...,M-1$ (the explicit time-dependence of the parameters has been dropped in the following),
 \begin{widetext}
 	\begin{equation}
 	\label{macierz}
 	Y^{ij}_{n}=\text{det} \! \left(
 	\begin{array}{ccccc}
 	|c_i|^2p_0&\cdots&0&c_i^*c_jp_ny^*_{n0}\\
 	\vdots&\ddots&\vdots&\vdots\\
 	0&\cdots&|c_i|^2p_{M-1}&c_i^*c_jp_ny^*_{nN-1}\\
 	c_ic_j^{*}p_ny_{n0}
 	&\cdots&c_ic_j^{*}p_ny_{nM-1}&|c_j|^2\sum_kp_k|y_{kn}|^2
 	\end{array}
 	\right)=|c_i|^{2M}|c_j|^2
 	\left[\prod_qp_q\sum_kp_k|y_{kn}|^2
 	-\sum_k\prod_{q\neq k}p_qp_n^2|y_{nk}|^2\right].
 	\end{equation}
 \end{widetext}
 This class of minors is of exactly the same form as in Ref.~\cite{roszak15a} and their analysis leads to an 
 analogous conclusion. There are three relevant situations:
 \begin{enumerate}
 	\item
 If all of the eigenvalues of the matrix $\hat{R}_{ii}(t)$, $p_k$,
 are non-zero, then
 \begin{equation}
 \label{nonzero}
 Y^{ij}_{n}=|c_i|^{2M}|c_j|^2\prod_qp_q\sum_k
 \left[p_k|y_{kn}|^2
 -\frac{p_n^2}{p_k}|y_{nk}|^2\right].
 \end{equation}
 If we now choose such $n'$ that the corresponding $p_{n'}$ is largest ($p_{n'}\ge p_k$ for all $k$),
 it is easy to show that $\sum_kp_k|y_{kn'}|^2\le p_{n'}$ and 
 $\sum_k\frac{p_{n'}^2}{p_k}|y_{n'k}|^2\ge p_{n'}$, since $\sum_k|y_{n'k}|^2=\sum_k|y_{kn'}|^2=1$ (because the operators
 $\hat{W}_{ij}(t)$ are unitary). 
 Hence, the minor $Y^{ij}_{n'}$ is equal to zero only, if for all $k$
 either $p_k=p_{n'}$ or $|y_{n'k}|=|y_{kn'}|=0$
 (when $\sum_kp_k|y_{kn'}|^2=p_{n'}$ and 
 $\sum_k\frac{p_{n'}^2}{p_k}|y_{n'k}|^2= p_{n'}$). Otherwise it is negative
 and there must be SEE present.
 
 In the situation, when $Y^{ij}_{n'}=0$, $n''$ should be studied, for which
 the eigenvalue $p_{n''}$ is second largest ($p_{n''}\le p_{n'}$
 and $p_{n''}\ge p_k$ for all $k\neq n'$). Both conditions stemming from
 $Y^{ij}_{n'}=0$ lead to the conclusion that 
 $\sum_kp_k|y_{kn''}|^2\le p_{n''}$ and 
 $\sum_k\frac{p_{n''}^2}{p_k}|y_{n'k}|^2\ge p_{n''}$.
 Hence, $Y^{ij}_{n'i}=0$ if and only if either $p_k=p_{n''}$ or $|y_{n''k}|=|y_{kn''}|=0$ for all $k$, otherwise it is negative.
 
 Repeating this reasoning for all minors in the order of diminishing $p_n$
 leads to the conclusion that there is SEE
 unless for all $k$ and $q$ either $p_k=p_q$ or $|y_{kq}|=|y_{qk}|=0$
 (and all principal minors from the class $Y^{ij}_{n}=0$).
 This is equivalent to the statement that entanglement has been generated
 unless $\left[\hat{R}_{ii}(t),\hat{W}_{ij}(t)\right]=0$.
 
 \item
 If only
 one of the eigenvalues of the matrix $\hat{R}_{ii}(t)$ is equal to zero
 (let us denote the corresponding eigenstate as $|r(t)\rangle$, so
 $p_r=0$), then 
 \begin{equation}
 Y^{ij}_{n}=-|c_i|^{2M}|c_j|^2
 \left(\prod_{q\neq r}p_q\right)p^2_n|y_{nr}|^2
 \end{equation}
 for $n\neq r$ and $Y^{ij}_{r}=0$. Hence, if there exists $|y_{nr}|\neq 0$
 for any $n\neq r$, then there is SEE.
 Otherwise the environmental state $|r(t)\rangle$ does not take part in 
 the system-environment evolution and does not need to be taken into 
 account, so entanglement generation may be probed using the minors
 of eq.~(\ref{nonzero}) after eliminating the state $|r(t)\rangle$
 from the subspace of the environment.
 
 Note that this situation can still be described using the commutation
 of $\hat{R}_{ii}(t)$ and $\hat{W}_{ij}(t)$.
 This is because, if there exists $|y_{nr}|\neq 0$, with $p_r=0$,
 but $p_n\neq 0$, then $Y^{ij}_{n}< 0$ and furthermore the
 commutation of $\hat{R}_{ii}(t)$ and $\hat{W}_{ij}(t)$ is impossible. Otherwise, for $|y_{nr}|= 0$ for all $n$ and there is entanglement in 
 the system unless $\hat{R}_{ii}(t)$ and $\hat{W}_{ij}(t)$ can be diagonalized
 in the same basis. Hence, if the condition $\left[\hat{R}_{ii}(t),\hat{W}_{ij}(t)\right]=0$ is not satisfied,
 then the system is entangled with its environment.
 
 \item
 If more than one eigenvalue of of the matrix $\hat{R}_{ii}(t)$ is equal to zero, then the class of minors given by eq.~(\ref{macierz}) is not
 a good class for the study of entanglement generation, since all of
 the minors in this class are always equal to zero.
 If for any of the states $\{|p\rangle\}$ for which $p_p=0$, we have
 $|y_{np}|=0$ for all $n$, then these 
 states can be elminated from the subspace of environmental states
 for the analysis of entanglement generation,
 since they do not take part in the system-environment evolution.
 If none, or only one relevant $|p\rangle$ state is present in the system,
 then the system should be treated as described above.
 
 Otherwise, if after eliminating the parts of the Hilbert space of the
 environment, which do not take part in the evolution, there are still
 $K\ge 2$ states from the subspace $\{|p\rangle\}$, then the analysis
 of entanglement generation requires a different class of principal minors. This is obtained
 by crossing out all but one of the rows and columns corresponding to diagonal elements
 equal to zero in the matrix given on the left in eq.~(\ref{macierz}),
 under the determinant. This new class has to be described by two indices,
 where the new index $r$ denotes the single state with $p_r=0$ taken 
 into account, while the scope of index $n$ does not encompass 
 states from the susbspace $\{|p\rangle\}$. Hence, we have
 a new class of minors
 \begin{equation}
 \tilde{Y}_{nr}^{ij}=-|c_i|^{2(M-K+1)}|c_j|^2
 \left(\prod_{q\notin \{p\}}p_q\right)p^2_n|y_{nr}|^2,
 \end{equation}
 where $\{p\}$ denotes the set of indices 
 for which $p_p=0$. 
 The minors $\tilde{Y}_{nr}^{ij}$ are negative when $|y_{nr}|\neq0$,
 so in the described situation, a negative minor must exist
 and entanglement is present. Furthermore, since the states corresponding
 to the non-zero element of the $\hat{W}_{ij}(t)$ operator, $|y_{nr}|$,
 have different occupations in the environmental state $\hat{R}_{ii}(t)$,
 $p_n\neq p_r=0$, $\hat{W}_{ij}(t)$ and $\hat{R}_{ii}(t)$ do not commute.
 \end{enumerate}
 
 Hence, we have shown that there is SEE 
 present in the density matrix (\ref{sigma}) of a system of any size $N$
 and environment of any size $M$, if for any $i\neq j$, where $i$ and $j$
 describe states of the system, the environmental density matrix
 $\hat{R}_{ii}(t)$ does not commute with the operator $\hat{W}_{ij}(t)$,
 so the conditions (\ref{warNequiv}) are broken. These conditions are
 equivalent to the qubit-like conditions (\ref{warN}), so entanglement
 is present in state (\ref{sigma}), if for any $i\neq j$ we have
 $\hat{R}_{ii}(t)\neq \hat{R}_{jj}(t)$.
 
 \section{\label{appB} Proof of entanglement generation when any 
 	of the additional conditions are broken,
 	while the qubit-like conditions are satisfied}
 
 We will continue to study a system of arbitrary size $N$ coupled
 via pure-dephasing interaction to an environment of again arbitrary
 size $M$, but now we assume that all of the qubit-like conditions
 (\ref{warN}), which are necessary for system-environment separability
 (but not sufficient), are satisfied. Hence, the system-environment density matrix
 can be written as
 \begin{eqnarray}
 \hat{\sigma}(t)&=&\sum_{i}|c_i|^2|i\rangle\langle i|\hat{R}_{00}(t)\\
 \nonumber
 &&+
 \sum_{i>j}\left(c_ic^*_j|i\rangle\langle j|\hat{R}_{00}(t)\hat{W}_{ij}(t)
 +\mathrm{H.c.}\right),
 \end{eqnarray}
 since (\ref{warN}) implies $\hat{R}_{ii}(t)=\hat{R}_{00}(t)$, for all $i$.
 After partial transposition with respect to the system, we get
 \begin{eqnarray}
 \label{transp}
 \hat{\sigma}^{T_S}(t)&=&\sum_{i}|c_i|^2|i\rangle\langle i|\hat{R}_{00}(t)\\
 \nonumber
 &&+
 \sum_{i>j}\left(c_ic^*_j|j\rangle\langle i|\hat{R}_{00}(t)\hat{W}_{ij}(t)
 +\mathrm{H.c.}\right).
 \end{eqnarray}
 
 As before, we will use the Peres-Horodecki criterion and the fact
 that a matrix has negative eigenvalues, if and only if at least one of
 its principal minors is negative. Hence, the existence of a negative
 principal minor of the density matrix after partial transposition (\ref{transp}) means
 that SEE has been generated
 at time $t$ during the evolution.
 
 Since the desired set of separability conditions (\ref{wardodN})
 is qualitatively different than in Appendix \ref{appA}, so will the
 studied set of principal minors be. We start by choosing three system
 states $i$, $j$, and $l$ and symmetrically 
 eliminating all rows and columns from
 the matrix (\ref{transp}) which describe the system-environment
 occupations and coherences not confined to the
 $\{|i\rangle,|j\rangle,|l\rangle\}$ subspace of the system. This
 yields the matrix (the explicit time dependence is omitted further on)
 \begin{equation}
 \label{Mijl}
 \hat{M}_{ijl}=\left(
 \begin{array}{ccc}
 |c_i|^2\hat{R}_{00}&c_ic^*_j\hat{W}_{ji}\hat{R}_{00}&
 c_ic^*_l\hat{W}_{li}\hat{R}_{00}\\
 c^*_ic_j\hat{R}_{00}\hat{W}^{\dagger}_{ji}&|c_j|^2\hat{R}_{00}&
 c_jc^*_l\hat{W}_{lj}\hat{R}_{00}\\
 c^*_ic_l\hat{R}_{00}\hat{W}^{\dagger}_{li}
 &c^*_jc_l\hat{R}_{00}\hat{W}^{\dagger}_{lj}
 &|c_l|^2\hat{R}_{00}
 \end{array}
 \right).
 \end{equation}
 
 It is now convenient to write the matrix (\ref{Mijl}) in terms 
 of eigenstates which diagonalize both $\hat{R}_{00}$ and $\hat{W}_{ji}$,
 which we denote as $|n\rangle$.
 Note, that the basis which diagonalizes $\hat{W}_{li}$
 (or $\hat{W}_{lj}$) also diagonalizes 
 $\hat{R}_{00}$, but not necessarily $\hat{W}_{ji}$. In this basis
 we have
 \begin{subequations}
 \begin{eqnarray}
 \hat{R}_{00}&=&\sum_np_n|n\rangle\langle n|,\\
 \hat{W}_{ji}&=&\sum_ne^{i\phi_n}|n\rangle\langle n|,\\
 \hat{W}_{li}&=&\sum_{nm}x_{nm}|n\rangle\langle m|,\\
 \hat{W}_{lj}&=&\hat{W}_{li}\hat{W}^{\dagger}_{ji}
 =\sum_{nm}x_{nm}e^{-i\phi_m}|n\rangle\langle m|.
 \end{eqnarray}
 \end{subequations}
 In this basis the matrix $\hat{M}_{ijl}$ can be written in an identical
 form as the partially transposed qutrit-environment density matrix,
 which satisfies the qubit-like conditions (\ref{war3}),
 given by eq.~(\ref{transpose}),
 \begin{widetext}
 	\begin{equation}
 	\label{transpose1}
 	\hat{M}_{ijl}=\left(\begin{array}{ccc}
 	|c_i|^2\sum_n p_n|n\rangle\langle n|
 	&c_ic_j^*\sum_n p_ne^{i\phi_n}|n\rangle\langle n|
 	&c_lc_i^*\sum_{nm} p_mx_{nm}|n\rangle\langle m|\\
 	c_jc_i^*\sum_n p_ne^{-i\phi_n}|n\rangle\langle n|
 	&|c_j|^2\sum_n p_n|n\rangle\langle n|
 	&c_lc_j^*\sum_{nm} p_me^{-i\phi_m}x_{nm}|n\rangle\langle m|\\
 	c_ic_l^*\sum_{nm} p_nx_{mn}^*|n\rangle\langle m|
 	&c_jc_l^*\sum_{nm} p_ne^{i\phi_n}x_{mn}^*|n\rangle\langle m|
 	&|c_l^2\sum_n p_n|n\rangle\langle n|\\
 	\end{array}
 	\right).
 	\end{equation}
 \end{widetext}
 
The class of principal minors for the study
 of qubit-environment entanglement generation, when the conditions
 (\ref{warN}) are satisfied is also the same as in Sec.~(\ref{sec3c}). They are determinants of 
 $3\times 3$ matrices obtained by symmetrically crossing out all but 
 one row and column, the diagonal elements of which are proportional
 to $|c_i|^2$, $|c_j|^2$, and $|c_l|^2$, respectively.
 Furthermore, the two rows and columns left over with diagonal elements
 proportional to $|c_i|^2$ and $|c_j|^2$ correspond to the $k$-th
 environmental state (the diagonal element is proportional to $p_k$), while
 the row and column left over with a diagonal element proportional to
 $|c_l|^2$ corresponds to the $q$-th state of the environment
 (the diagonal element is proportional to $p_q$).
 The minors of interest are therefore labeled by to environmental
 indices $k$ and $q$, and are given by
 \begin{widetext}
 	\begin{equation}
 	X^{ijl}_{kq}=\det\left(
 	\begin{array}{ccc}
 	|c_i|^2p_k&c_ic_j^*p_ke^{i\phi_k(t)}&c_lc_i^*p_qx_{kq}\\
 	c_i^*c_jp_ke^{-i\phi_k(t)}&|c_j|^2p_k&c_lc_j^*p_qe^{-i\phi_q(t)}x_{kq}\\
 	c_2^*c_lp_qx_{kq}^*&c_l^*c_jp_qe^{i\phi_q(t)}x_{kq}^*&|c_l|^2p_q
 	\end{array}
 	\right)=-2|c_ic_jc_l|^2p_k^3|x_{kq}|^2\left[1-\cos\left(
 	\phi_k(t)-\phi_q(t)\right)\right],
 	\end{equation}
 \end{widetext}
 since $\left[\hat{R}_{00},\hat{W}_{li}\right]=0$ implies that 
 for $x_{kq}\neq 0$, we must have $p_k=p_q$.
 No minor $X^{ijl}_{kq}$ can be positive, and they can be equal to zero only
 in four situations, two of which are trivial. The first trivial one
 is when
 at least one of the initial qubit occupations is zero and 
 therefore $|c_ic_jc_l|^2=0$,
 so that the studied system state is of lower dimension than $N$
 and the system-environment density matrix should be adjusted accordingly. In the second trivial situation 
 $p_k=p_q=0$.
 Because for the studied state the qubit like conditions (\ref{warN})
 are satisfied, it follows that for $p_k=p_q=0$, $x_{kn}=0$ and
 $x_{qn}=0$ for all $n$ with $p_n\neq 0$, and $|k\rangle $ and $|q\rangle$ can be eliminated
 from subspace of environmental degrees of freedom taken into account.
 
 Otherwise, the principal minor $X^{ijl}_{kq}$ is equal to zero 
 (is non-negative) only, if
either $x_{kq}=0$ or
 $e^{i\phi_k(t)}-e^{i\phi_q(t)}=0$.
 If all of the principal minors of this class, $X^{ijl}_{kq}$,
 for fixed $i$, $j$, and $l$, but for every possible value of $k\neq q$
 are non-negative, then it is equivalent to write
 \begin{equation}
 \label{wardod2N}
 \left[\hat{W}_{ji}(t),\hat{W}_{li}(t)\right]=0.
 \end{equation} 
 Obviously $X^{ijl}_{kq}<0$ for any $k$ or $q$ means
 that there is qubit-environment entanglement in the state, so
 if the condition (\ref{wardod2N}) is not met, there must be entanglement 
 in the system.
 
 Repeating the procedure detailed above for all $i\neq j\neq l$ yields a set of conditions of the form of eq.~(\ref{wardod2N}) for system-environment
 states which already fulfill all of the conditions (\ref{warN}).
 If for any system states $i\neq j\neq l$, the condition (\ref{wardod2N}) is not
 satisfied (for at least one pair of environment states $k$ and $q$), then there is entanglement between the system and the
 environment in the state. Otherwise, for all $i\neq j\neq l$, operators
  $\hat{W}_{ji}(t)$ and $\hat{W}^{\dagger}_{li}(t)$ commute.
  This is enough to show that there is no SEE,
  if the system is a qutrit, since in this case 
  $\hat{R}_{00}$, $\hat{W}_{01}(t)$, $\hat{W}_{12}(t)$, and $\hat{W}_{20}(t)$
  have a common eigenbasis
  and the system-environment density matrix (\ref{sigma4}) can be written
  in a separable form.
  For larger systems, for there
  to exist an environmental basis, which diagonalizes
  $\hat{R}_{00}(t)$ and all operators $\hat{W}_{ij}(t)$,
  also $\hat{W}_{ij}(t)$ and $\hat{W}_{kl}(t)$ must commute
  for all $i\neq j\neq k\neq l$ (in the case of the qutrit, such four system indices do not exist). Incidentally, no new class of minors
  is needed in this case, since the commutation of two $\hat{W}_{ij}(t)$
  operators described by four different indices can be derived from
  commutation of $\hat{W}_{ij}(t)$
  operators with only three different indices. If the condition
  (\ref{wardod2N}) is fulfilled for all $i\neq j\neq l$, we get
  \begin{eqnarray}
  \nonumber
  \left[\hat{W}_{ij}(t),\hat{W}_{kl}(t)\right]&=&
  \hat{W}_{ij}(t)\hat{W}_{kl}(t)-\hat{W}_{kl}(t)\hat{W}_{ij}(t)\\
  \nonumber
  &=&\hat{W}_{ij}(t)\hat{W}_{kj}(t)\hat{W}_{jl}(t)-
  \hat{W}_{kl}(t)\hat{W}_{ij}(t)\\
  \nonumber
  &=&\hat{W}_{kj}(t)\hat{W}_{ij}(t)\hat{W}_{jl}(t)-
  \hat{W}_{kl}(t)\hat{W}_{ij}(t)\\
  \nonumber
  &=&\hat{W}_{kj}(t)\hat{W}_{jl}(t)\hat{W}_{ij}(t)-
  \hat{W}_{kl}(t)\hat{W}_{ij}(t)\\
  \nonumber
  &=&\hat{W}_{kl}(t)\hat{W}_{ij}(t)-
  \hat{W}_{kl}(t)\hat{W}_{ij}(t)=0.
  \end{eqnarray}
  Here we used the fact that the condition (\ref{wardod2N}) also means that
  \begin{eqnarray}
  \nonumber
  \left[\hat{W}_{ij}(t),\hat{W}_{il}(t)\right]&=&0,\\
  \nonumber
  \left[\hat{W}_{ji}(t),\hat{W}_{il}(t)\right]&=&0,\\
  \nonumber
  \left[\hat{W}_{ji}(t),\hat{W}_{li}(t)\right]&=&0,
  \end{eqnarray}
  since $\hat{W}_{ji}(t)=\hat{W}^{\dagger}_{ij}(t)$
  and the operators $\hat{W}_{ji}(t)$ are unitary,
  and the fact that $\hat{W}_{kl}(t)=\hat{W}_{kj}(t)\hat{W}_{jl}(t)$.
  
  Hence, if for all $i\neq j\neq l$, the condition (\ref{wardod2N})
  is fulfilled, then all operators $\hat{W}_{ij}(t)$ commute
  with each other.
  This means that the system-environment density matrix may be 
  written in a basis $\{|n(t)\rangle\}$, which is not only the eigenbasis of
  $\hat{R}_{00}(t)$, but also diagonalizes all 
  of the operators $\hat{W}_{ij}(t)$ (and the initial, pointer basis of the system)
  in the obviously separable form
  \begin{equation}
  \label{sigmasepN}
  \hat{\sigma}(t)=\sum_n p_n(t)\hat{\rho}_n(t)\otimes |n(t)\rangle\langle n(t)|,
  \end{equation}
  where the density matrices of the system conditional on the
  state of the environment are given by
  \begin{equation}
  \hat{\rho}_n(t)=\sum_{ij}c_ic^*_je^{i\phi_n^{ij}(t)}|i\rangle\langle j|,
  \end{equation}
  with $|i\rangle$ and $|j\rangle$ denoting the system states for which
  the Hamiltonian (\ref{ham}) is diagonal, and the oscillating factors
  being a product of the diagonalization of the operators
  \begin{equation}
  \hat{W}_{ij}(t)=\sum_ne^{i\phi_n^{ij}(t)}|n(t)\rangle\langle n(t)|,
  \end{equation}
  $e^{i\phi_n^{ii}(t)}=1$, and $e^{i\phi_n^{ji}(t)}=e^{-i\phi_n^{ij}(t)}$.
 
 In a last step, let us reduce the number of relevant conditions
 (\ref{wardod2N}), by restricting ourselves to the independent ones,
 which cannot be derived from one another.
 This involves choosing a fixed index $i$ in the set of conditions
 (\ref{wardod2N}), say $i=0$. The set of all independent conditions is
 such that for all $j>l> 0$,
 \begin{equation}
 \label{wardod2Nost}
 \left[\hat{W}_{j0}(t),\hat{W}_{l0}(t)\right]=0.
 \end{equation} 
 All other commutation relations, $\left[\hat{W}_{ij}(t),\hat{W}_{kl}(t)\right]=0$, for any $i,j,k,l$,
 can be derived from (\ref{wardod2Nost}) using $\hat{W}_{ij}(t)=\hat{W}_{i0}(t)\hat{W}_{0j}(t)$, $\hat{W}_{ii}(t)=\unit$,
 and the fact that the operators $\hat{W}_{ij}(t)$ are unitary, so
 commutation is unaffected by 
 hermitian conjugation of one or both operators. 
 Hence, there are $(N-1)(N-2)/2$ independent commutation relations
 for a system of size $N$,
 the fulfillment of which guarantees separability of the system
 from its environment, if system-environment state 
 already satisfies the $(N-1)$ qubit-like conditions (\ref{warN}).
 
 \end{appendix}

\end{document}